\documentclass{WileyMSP-template}
\usepackage{graphicx}
\usepackage{amsmath}
\usepackage{dcolumn}
\usepackage{bm}
\usepackage{cite}

\begin{document}

\pagestyle{fancy}
\rhead{\includegraphics[width=2.5cm]{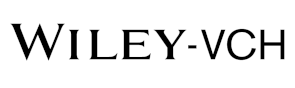}}

\title{Predicting the Structure and Stability of Oxide Nanoscrolls from Dichalcogenide Precursors}

\maketitle


\author{Adway Gupta}
\author{Arunima K. Singh}


\begin{affiliations}
Adway Gupta, Dr. Arunima K. Singh\\
Department of Physics, Arizona State University, Tempe, Arizona  85287-1504\\
Email Address: arunimasingh@asu.edu
\end{affiliations}



\begin{abstract}
Low-dimensional nanostructures such as nanotubes, nanoscrolls, and nanofilms have found applications in a wide variety of fields such as photocatalysis, sensing, and drug delivery. Recently, Chu \textit{et al}\cite{Chu2017} demonstrated that nanoscrolls of Mo and W transition metal oxides, which do not exhibit van der Waals (vdW) layering in their bulk counterparts, can be successfully synthesized using a plasma processing of corresponding layered transition metal dichalcogenides. In this work, we employ data mining, first-principles simulations, and physio-mechanical models to theoretically examine the potential of other dichalcogenide precursors to form oxide nanoscrolls. Through data mining of bulk and two-dimensional materials databases, we first identify dichalcogenides that would be mostly amenable to plasma processing on the basis of their vdW layering and thermodynamic stability. To determine the propensity of forming a nanoscroll, we develop a first-principles simulation-based physio-mechanical model to determine the thermodynamic stability of nanoscrolling as well as the equilibrium structure of the nanoscrolls, i.e. their inner radius, outer radius, and interlayer spacing. We validate this model using the experimental observations of Chu \textit{et al}'s study and find an excellent agreement for the equilibrium nanoscroll structure. Furthermore, we demonstrate that the model's energies can be utilized for a generalized quantitative categorization of nanoscroll stability. We apply the model to study the oxide nanoscroll formation in MoS$_2$, WS$_2$, MoSe$_2$, WSe$_2$, PdS$_2$, HfS$_2$ and GeS$_2$, paving the way for a systematic study of oxide nanoscroll formation atop other dichalcogenide substrates. 
\end{abstract}


\section{Introduction}
Low-dimensional materials, such as two-dimensional (2D) materials, nanowires, quantum dots, and nanoscrolls have garnered significant attention due to their unique properties which result from their reduced dimensionality and quantum confinement. Among low-dimensional materials, nanoscrolls present new opportunities for efficient energy storage, photocatalysis,\cite{tmdctmo,tmocatalysis1,tmocatalysis2}, nanoelectronics \cite{cnshydrogen,Mos2nsphoto,Sayyad2023,gupta2022anomalous,guo2021exfoliation,li2019anomalous} and drug delivery. A nanoscroll is a rolled-up sheet of nanoribbon or a 2D monolayer. Such nanoribbons or monolayers can be extracted by cleaving bulk materials along planes that have weak van der Waals (vdW) bonding. Among well-studied nanoscroll materials are those of transition metal dichalcogenides (TMDCs), which exhibit layered bulk structures with well-defined directions of vdW bonding.

\begin{figure}[h]
    \centering
    \includegraphics[width=4in]{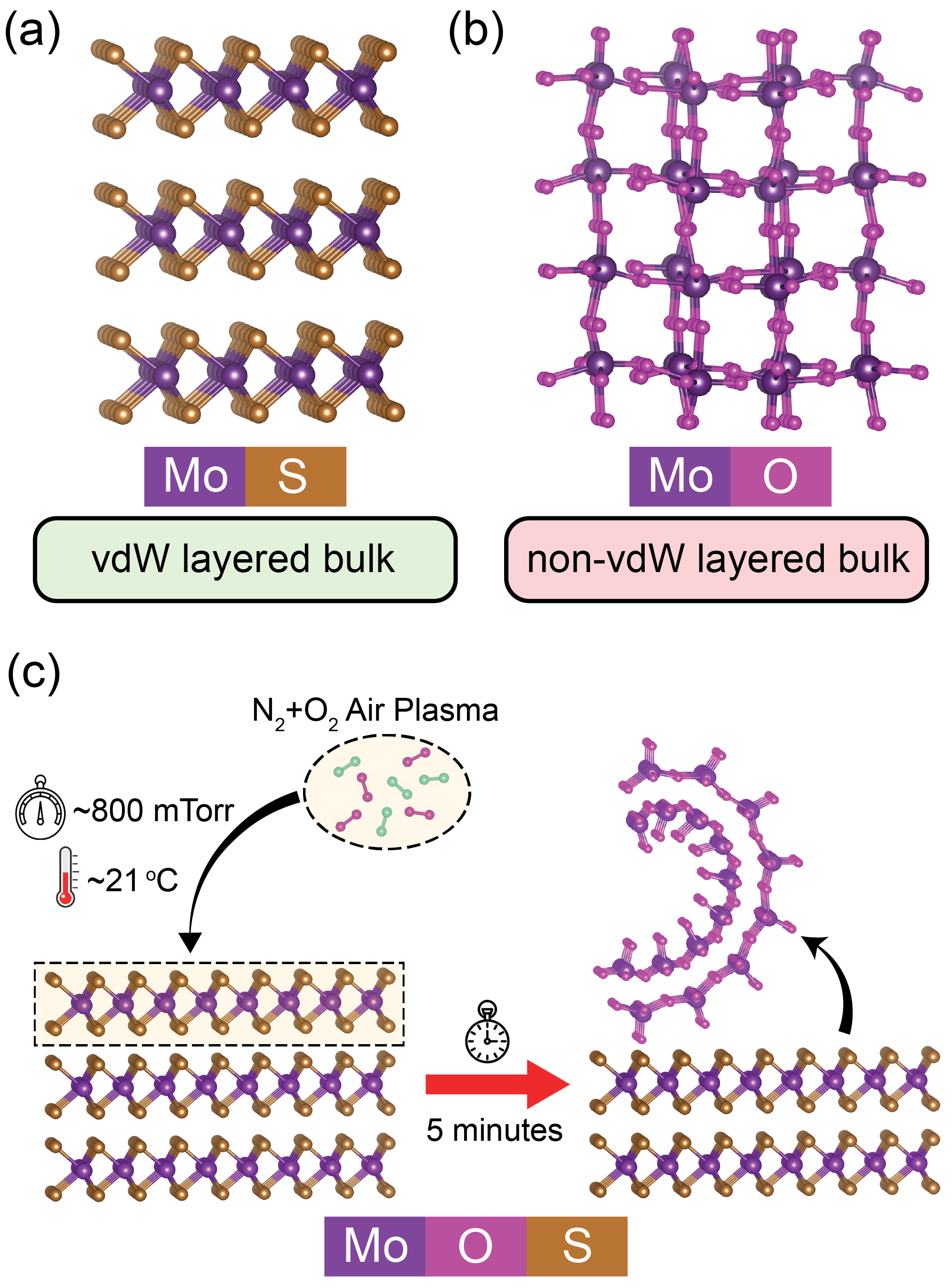}
    \caption{A schematic of the bulk crystal structures of, (a) MoS$_2$ and the corresponding bulk oxide, (b) MoO$_3$ (c) A schematic of formation of oxide nanoscrolls from layered dichalcogenide bulk materials via the plasma assisted conversion synthesis (PACS). }
    \label{PACS_vdw}
\end{figure}

Recently, nanoscrolls of materials that do not exhibit vdW bonding in the bulk were synthesized by Chu \textit{et al}\cite{Chu2017}. They obtained nanoscrolls of transition-metal oxides (TMOs) that do not have a vdW layered bulk counterpart by performing a plasma assisted conversion synthesis (PACS) process on layered bulk TMDCs. The process involved placing the TMDCs in a plasma chamber with 40-80 mTorr pressure and exposing them to atmospheric air plasma for 5 minutes to 800 mTorr pressure. They observed that the bulk MoS$_2$ and bulk WS$_2$ TMDCs converted to respective TMO nanoscrolls. Specifically, MoS$_2$ was converted to MoO$_3$ nanoscrolls and WS$_2$ was converted to WO$_3$ nanoscrolls, as characterized by their XPS, AFM, and EDS measurements. Figure \ref{PACS_vdw}(a) and \ref{PACS_vdw}(b) show  the crystal structures of bulk MoS$_2$ and MoO$_3$, respectively. Chu \textit{et al}, however, did not obtain any nanoscrolls from the PACS process for MoSe$_2$ and WSe$_2$, and instead only obtained nanofilms of their oxides. Figure \ref{PACS_vdw}(c) shows a schematic of the layer-by-layer conversion of the TMDCs to TMOs in the PACS process. 

In this article we employ data-mining, first-principles simulations, and physio-mechanical models to theoretically determine the propensity of a dichalcogenide to form oxide nanoscrolls. To determine the propensity of forming a nanoscroll, we develop a first-principles simulation-based physio-mechanical model that includes various driving forces for the formation of a nanoscroll, including bending energies, interlayer surface energies, and layer-substrate interaction energies. We then utilize this model to determine the thermodynamic stability of nanoscrolling as well as the equilibrium structure of the nanoscrolls, i.e. their inner radius, outer radius, and interlayer spacing. We validate this model against Chu \textit{et al.}'s work and find that the model gives an excellent agreement with the experimentally reported structures. In addition, we find that the model energies can be utilized for quantitative categorization of the stability of the nanoscrolls. We data-mined other dichalcogenide candidate materials from bulk and two-dimensional materials databases that would be mostly amenable to the plasma processing. These include MoS$_2$, MoSe$_2$, WS$_2$, WSe$_2$,  PdS$_2$, HfS$_2$, and GeS$_2$. We model their oxide nanoscrolls' thermodynamic stability and equilibrium structures. We find that the often ignored layer-substrate interaction energies' effect on the thermodynamical stabilities of the oxide nanoscrolls is significant but on their equilibrium structures is nominal. Our work paves the way for a systematic, predictive and mechanistic study of oxide nanoscroll formation from dichalcogenide precursors, guiding their economical and large-scale experimental realization and applications.

\section{Results and Discussion}

\subsection{Physio-Mechanical Model of Nanoscrolls}\label{model}

\begin{table*}[ht]
    \centering
    \begin{tabular}{|c|c|c|c|c|c|}
         \hline
         Precursor & Selected & Oxide  & Oxide in-plane  & Oxide  & Experimentally synthesized   \\
          & Oxide & spacegroup & lattice parameters & source & oxide monolayer? \\
         \hline
         MoS$_2$,MoSe$_2$ & MoO$_3$ & P$2_1$/m & $a$ = 3.71~\AA ; $b$ = 3.90~\AA & Haastrup \textit{et al}\cite{haastrup2018computational} & Yes \\
         WS$_2$,WSe$_2$, & WO$_3$ & P$\bar{1}$ & $a$ = 3.74~\AA ; $b$ = 3.83~\AA & Negreiros \textit{et al}\cite{Negreiros2019} & Yes \\
         PdS$_2$ & PdO$_2$ & P$\bar{3}$m1 & $a$ = $b$ = 3.07~\AA & Haastrup \textit{et al}\cite{haastrup2018computational} & No \\
         HfS$_2$ & HfO$_2$ & P$\bar{3}$m1 & $a$ = $b$ = 3.24~\AA & Weng \textit{et al}\cite{weng2018honeycomb} & No \\
         GeS$_2$ & 1T-GeO$_2$  & P$\bar{3}$m1, & $a$ = $b$ = 2.92~\AA , & Haastrup \textit{et al}\cite{haastrup2018computational}  & No  \\
         & m-GeO$_2$ & C2/m & $a$ = $b$ = 5.21~\AA & Singh \textit{et al}\cite{Singh2017} & No \\
         \hline
    \end{tabular}
    \caption{Table shows all the oxide-dichalcogenide pairs that are considered in this study along with the chemical formula, lattice constants, space groups, and the state of experimental synthesis for the 2D form of the oxide.}
    \label{selected_TMO}
\end{table*}

\begin{figure}
    \centering
    \includegraphics[width=4in]{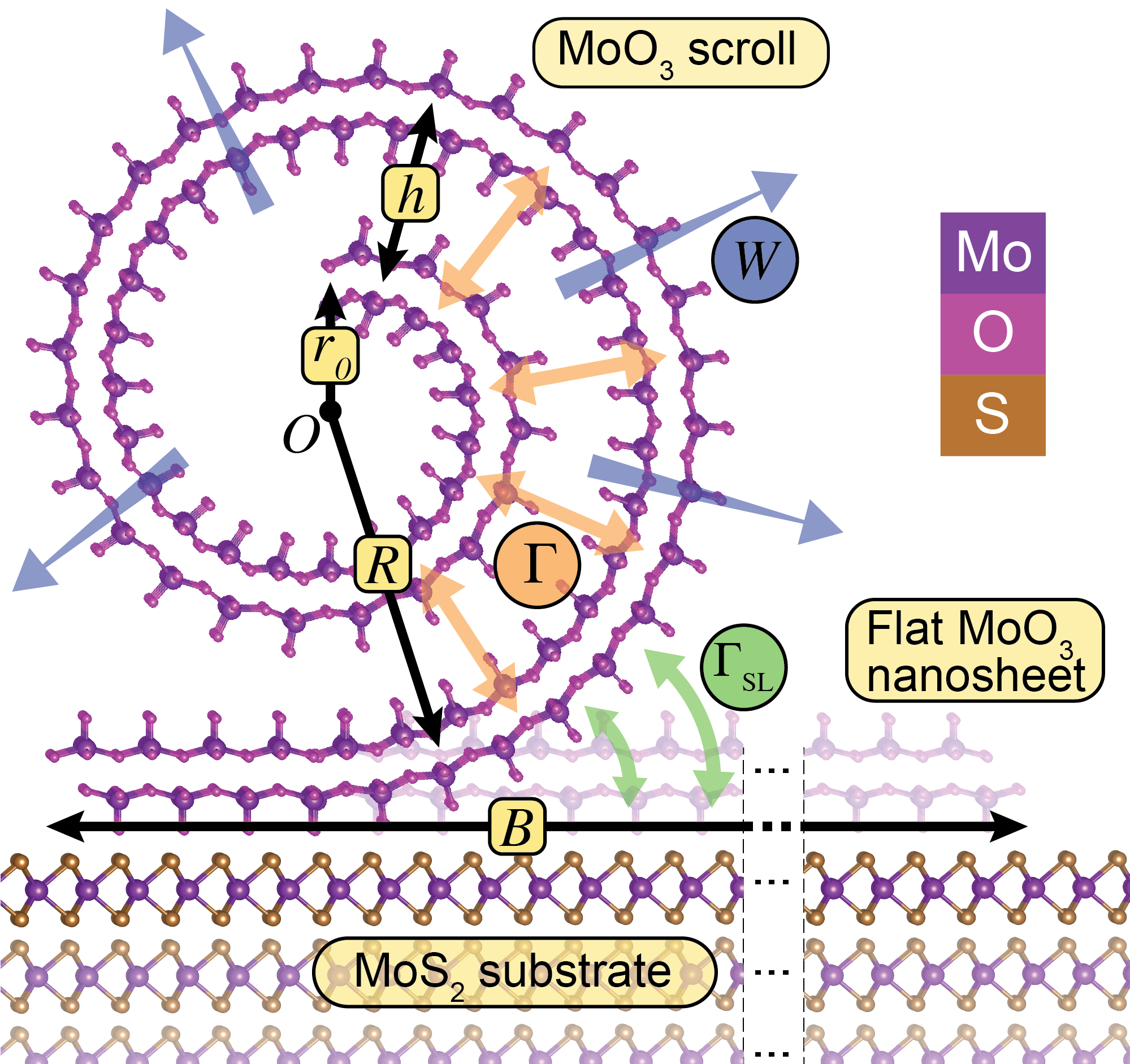}
    \caption{A schematic of an MoO$_3$ nanoscroll formed atop an MoS$_2$ precursor. The blue, orange, and green arrows indicate the bending ($W$), interlayer ($\Gamma$) and layer-substrate interaction ($\Gamma_{S-L}$) energies respectively acting on the nanoscroll as well as their directions. The nanoscroll's geometrical parameters, i.e., inner radius ($r_0$), outer radius ($R$), interlayer spacing ($h$), and length of the flat nanosheet ($B$), are also indicated.}
    \label{forces_fig}
\end{figure}

A nanoscroll structure can be uniquely defined by any three of its following four geometric parameters: its core radius, $r_0$, its outer radius, $R$, its interlayer spacing, $h$, and the length of the flat nanosheet, $B$, that is rolled to form the nanoscroll \cite{Shi2010,Shi2010_2}. These four parameters are depicted in Figure \ref{forces_fig}. The relation between the four parameters of a nanoscroll is as follows, \cite{Shi2010,Shi2010_2}

\begin{equation}\label{sheet_width}
    B = \frac{\pi}{h}(R^2-r_0^2) 
\end{equation}

A nanoscroll that is in thermodynamic equilibrium has a structure that minimizes its total energy. Thus, in order to determine an equilibrium scroll structure, we need to determine the value of the geometric parameters for which the total energy of the structure is minimized. Prior studies have shown that two distinct energies contribute to the total energy of a nanoscroll -- 1) the interlayer energy, $\Gamma$, and 2) the bending energy, $W$.\cite{Shi2010,Shi2010_2}

The interlayer energy, $\Gamma$ is the energy associated with the interaction between the surface of the successive layers of the scroll. This energy would be the same as the product of the surface energy density, $\gamma$, and the area for the case of thin films. However, for a nanoscroll, due to its geometry, the interlayer energy, $\Gamma$, is given by the following equation, \cite{Shi2010,Shi2010_2}

\begin{equation}\label{gammaeq}
    \Gamma = 2\pi\gamma\Big((R-r_0) + \frac{R^2-r_0^2}{2R}\Big)
\end{equation}

where $R$ and $r_0$ are the outer radius and core radius of the nanoscroll structure, and $\gamma$ is the interlayer surface energy per unit area of the nanoscroll. The effect of $\Gamma$ is depicted by orange arrows in Figure \ref{forces_fig}.

The second energy that has been shown to determine the stability of a nanoscroll is the bending energy, $W$. The effect of $W$ is illustrated with blue arrows in Figure \ref{forces_fig}. As expected, the $W$ depends on the bending stiffness, $D$, of the materials  and is given by the following equation, 
\begin{equation}\label{Weq}
    W = \frac{\pi D}{h}\ln \frac{R}{r_0}
\end{equation}
where  $R$ and $r_0$ are again the outer radius and core radius, $h$ is the interlayer spacing of the nanoscroll, and $D$ is the bending stiffness of the 2D material\cite{Shi2010}.

In our work, a third contribution to the total energy of the nanoscrolls needs to be considered. This energy would account for the energy associated with the delamination of the oxide layer that is formed atop the bulk-layered dichalcogenide. It should be noted that the impact of the substrate on the thermodynamic stability of a nanoscroll structure has not been considered in prior literature. 

Thus, we introduce another energy term, the layer-substrate interaction energy, $\Gamma_{S-L}$, whose effect is illustrated by green arrows in Figure \ref{forces_fig}. We define $\Gamma_{S-L}$ as, 
\begin{equation}\label{gammasleq}
    \Gamma_{S-L} = \gamma_{S-L} \times B 
\end{equation}
where again, $B$ is the width of the original nanosheet, while $\gamma_{S-L}$ is the layer-substrate energy per unit area. 

In summary, the total energy of the nanoscrolls, $E_{\text{total}}$, can be given by the sum of all the three energy contributions and can be written as,  
\begin{equation}\label{totalEeq}
    E_{\text{total}} = W + \Gamma + \Gamma_{S-L}
\end{equation}

In the following sections, we first employ this new \textit{ab initio} physio-mechanical model of Equation \ref{totalEeq} which accounts for substrate effects for the following three tasks, 

\begin{enumerate}
\item Determine the geometry of the oxide nanoscroll structures from the MoS$_2$ and WS$_2$ precursors and compare them to those obtained by Chu \textit{et al}
\item Explain why nanoscrolls are not obtained for the MoSe$_2$ and WSe$_2$ precursors in Chu \textit{et al}'s experiments
\item Determine if any other layered materials can be used as precursors for the successful growth of oxide nanoscrolls
\end{enumerate}

 For task 3, we identify materials that can act as precursors for forming oxide nanoscrolls via the PACS process by data mining of existing materials' databases and then apply the physio-mechanical model to establish the structure and stability of oxide nanoscrolls that can be formed from these precursors. We identify dichalcogenide materials where the cation is one of the transition metals-- Sc, Ti, V, Cr, Mn, Fe, Co, Cu, Y, Zr, Nb, Mo, Tc,  Ru, Rh, Pd, Ag, Hf, Ta, W, Re, Os, Ir, Pt, Au by querying the well known Materials Project (MP) database\cite{jain2013commentary}. We also include Ge dichalcogenides since they are known to have layered dichalcogenides\cite{GeS2synthesis}. This resulted in 772 materials of the form MX$_2$ where M is one of the cations listed above and X= S, Se, Te. To make the investigation of nanoscroll structure and stability computationally tractable, we first imposed a criteria- selecting only those precursors that do not have any corresponding layered bulk oxides, thus limiting the study to oxides that cannot produce nanoscrolls from their 2D films but can only be produced by alternate methods like the PACS process. This resulted in 117 materials.
 
 To further downselect, we only selected bulk precursor materials that were layered and have been experimentally synthesized. To identify materials that satisfied these criteria we searched the MP database\cite{jain2013commentary} for bulk dichalcogenide materials which exhibit vdW layering. Note that to determine if the materials were layered or not, we used the pymatgen code\cite{ong2013python}. Materials that are experimentally synthesized have an ICSD ID\cite{icsd}reported in the MP database. In this step several materials are eliminated, for example, Mo$_3$S$_2$ is eliminated from the screening process due to the lack of bulk vdW layering, while MoS$_2$ is selected. Total 73 potential dichalcogenides were identified through the overall screening process and their detailed information can be found in Table S1. Additional information from 2D materials databases\cite{haastrup2018computational} are used to reduce this list of candidates to 16 materials, which are detailed in Table S2. The complete screening process is also diagrammatically detailed in Figure S1. Amongst the screened materials, we select the following 8 materials to exemplify the application of the physio-mechanical nanoscroll model-- MoS$_2$, WS$_2$, MoSe$_2$, WSe$_2$, PdS$_2$, HfS$_2$ and GeS$_2$. Table \ref{selected_TMO} shows the corresponding oxide formula, their lattice constants, spacegroup, and if these monolayer structures are predicted from first-principles simulations or obtained from experiments. 
 
 The requisite experimental data and information for performing tasks 1 and 2 is available in Chu \textit{et al}'s work (as discussed in detail in section \ref{NSresults}). For all three tasks, the \textit{ab initio} methods utilized and the results obtained for the \textit{ab initio} computed quantities namely $D$, $\gamma$, and $\gamma_{S-L}$ are described in the next section.

\begin{figure}
    \centering
    \includegraphics[width=3.4in]{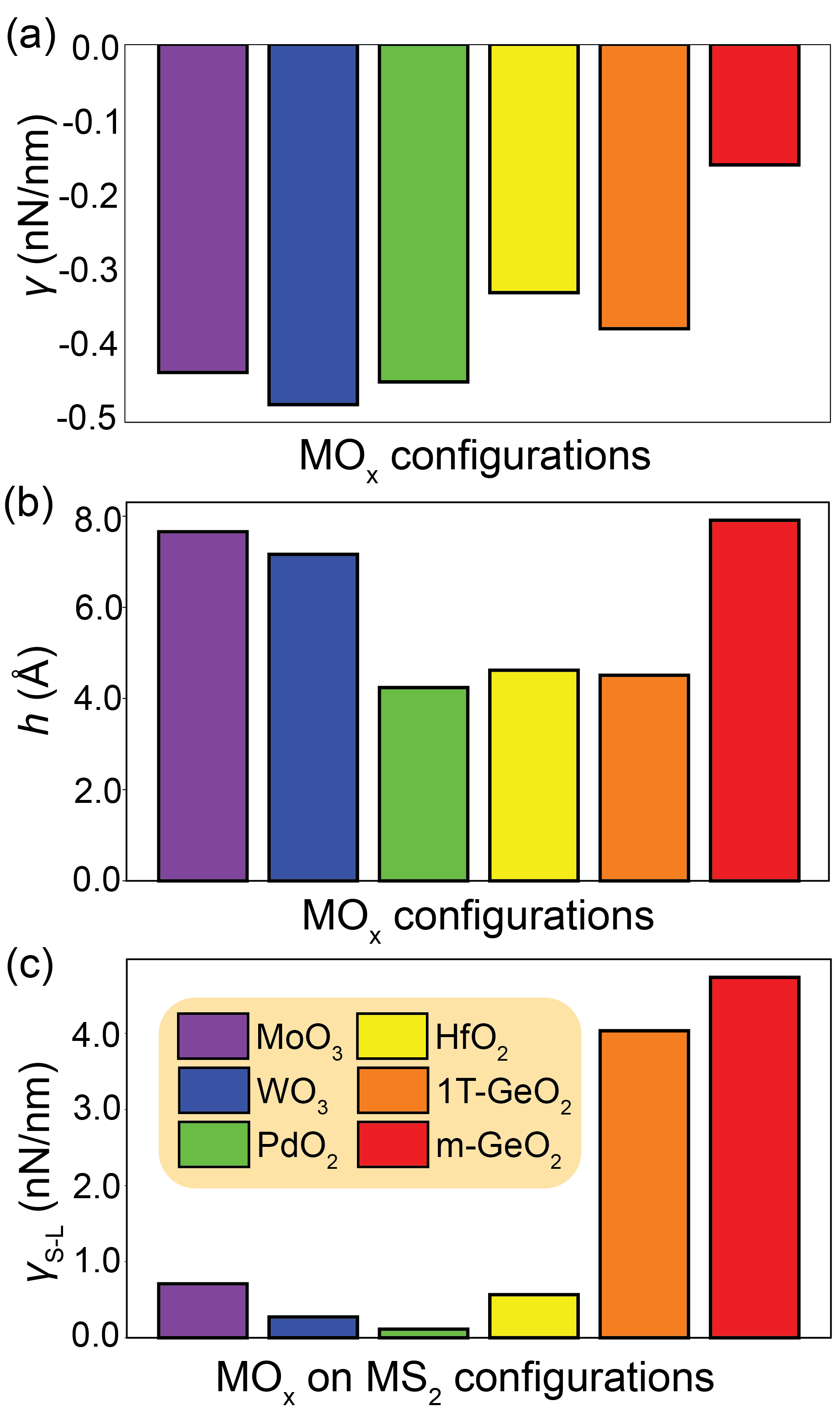}
    \caption{(a) shows the values of, $\gamma$, in nN/nm for all the selected oxides while (b) shows their equilibrium interlayer spacing, $h$, in \AA. (c) shows the corresponding, $\gamma_{S-L}$, in nN/nm  for all the oxide-disulphide pairs.}
    \label{gammasl_fig}
\end{figure}

\begin{figure*}
    \centering \includegraphics{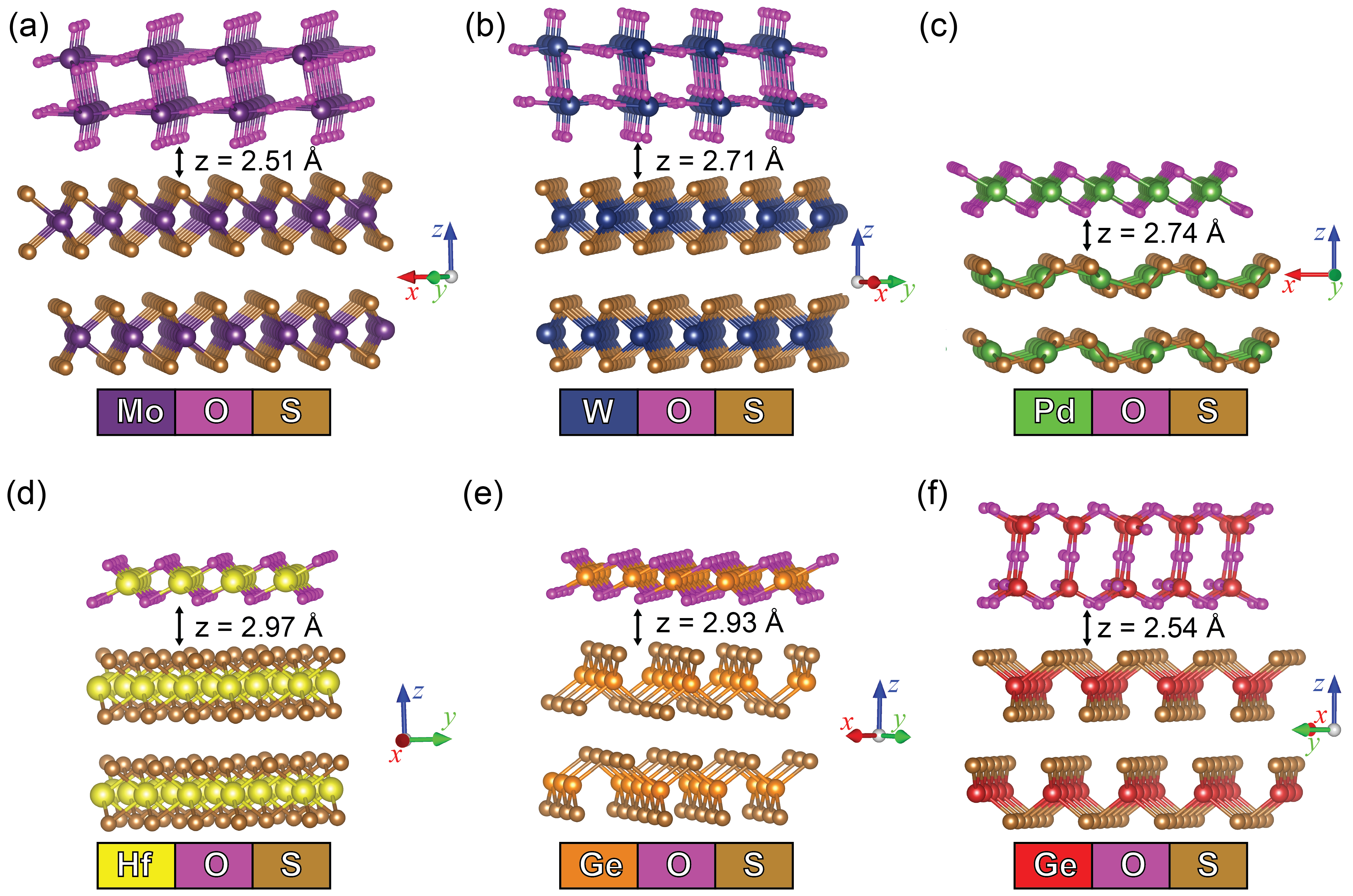}
    \caption{Figures show the oxides atop the corresponding disulphide heterostructures for (a) MoO$_3$-MoS$_2$ (b) WO$_3$-WS$_2$ (c) PdO$_2$-PdS$_2$ (d) HfO$_2$-HfS$_2$ (e) 1T-GeO$_2$-GeS$_2$ and (f) m-GeO$_2$-GeS$_2$. Numbers at the interface indicate the final z-separation of the heterostructures}
    \label{heterostructs}
\end{figure*}

\subsubsection{\textit{Ab initio} Calculation of Parameters for the Physio-Mechanical Nanoscroll Model}

To compute the equilibrium total energy and structure of a nanoscroll we employ the physio-mechanical model described in section \ref{model} with parameter values computed from the density-functional theory (DFT) first-principles simulation method. DFT with vdW corrections is known to provide excellent accuracy for the requisite parameters, in particular, the interlayer surface energy per unit area of the nanoscroll $\gamma$, the bending stiffness $D$, the layer-substrate interaction energy $\gamma_{S-L}$, and the interlayer spacing $h$. The remainder of the geometric parameters of a nanoscroll, namely $r_0$, $B$, and $R$ can be estimated by obtaining the minima of the total energy as a function of the core radii, $r_0$, assuming a given sheet size, $B$. The outer radius $R$ can then be obtained from Equation \ref{sheet_width}. Note that the variation of the equilibrium structure with varying $B$ can also be analyzed. In the following sections we describe in detail the vdW corrected DFT approach to calculate the required parameters, i.e., $\gamma$,  $D$, $\gamma_{S-L}$, and $h$.

\subsubsection{\textit{Ab initio} calculation of $\gamma$} \label{gamma}

We determine the interlayer interactions using \textit{ab initio} vdW corrected DFT calculations using the following equation, 
\begin{equation}\label{gamma_eq}
    \gamma = \frac{E_{\text{bulk}}-E_{\text{mono}}}{A}
\end{equation}
where $E_{\text{bulk}}$ is the energy of the layered bulk oxide, $E_{\text{mono}}$ is the energy of a monolayer oxide slab and $A$ is the surface area of these structures. The DFT calculations for the determination of $\gamma$ are based on the projector-augmented wave method as implemented in the plane-wave code VASP.\cite{Kresse1,Kresse2,Kresse3,Kresse4,Kresse5} All DFT simulations were performed using the vdW-DF-optB88 exchange-correlation functional,\cite{Klimes2011} with a cutoff energy for the plane wave basis as 520 eV and a $k$-grid density of 50 \AA$^{-3}$. The monolayers were simulated with a large vacuum spacing of 30~\AA~in order to eliminate effects from periodic images in the $z$-direction. The interlayer spacing of the oxide nanoscroll, $h$, were also determined from the relaxation of the layered bulk oxide. All structural relaxations were performed until the total forces are converged within $0.1$ eV/\AA~and the total energies are converged within $10^{-6}$ eV per unit cell.  

The $\gamma$ as well as the $h$, defined as the distance between the center of one layer to that of its periodic image in the $z$ direction for the layered bulk oxide structure, for all our focus materials are shown in Figure \ref{gammasl_fig}(a)-(b). We see that the values of $\gamma$ are fairly uniform across the selected oxides, except for m-GeO$_2$, indicating m-GeO$_2$ has a reduced tendency to form bilayers or multilayers. Also, the $h$ for PdO$_2$, HfO$_2$ and 1T-GeO$_2$ are smaller than those for the other materials as they are thinner hexagonal 1T-structures.

\subsubsection{\textit{Ab initio} calculation of $\gamma_{S-L}$} \label{gammaSL}

The layer-substrate interaction energy, $\gamma_{S-L}$, can be computed as, 
\begin{equation}
    \gamma_{S-L} = \frac{E_{\text{hetero}}-(E_{\text{2D}}^{\text{relax}}+E_{\text{sub}})}{A}
\label{eq7}
\end{equation}
where $E_{\text{hetero}}$ is the energy of the oxide monolayer placed on the dichalcogenide substrate slab and $A$ is the area in the $xy$ plane of this heterostructure. $E_{\text{sub}}$ is the energy of the dichalcogenide substrate slab and $E_{\text{2D}}^{\text{relax}}$ is the energy of the oxide monolayer slab. vdW-corrected DFT simulations are performed to obtain these energies. The non interacting oxide monolayer and the dichalcogenide substrate energies are computed in slab geometry with vacuum spacing of more than 30~\AA~ in the $z$-direction in order to eliminate the effects from periodic images. For the heterostructure simulations, the oxide monolayers and the top 2 layers of the substrates are allowed to relax, keeping the rest of the substrate atoms fixed. For the structural relaxations, the total forces are converged within $0.02$ eV/\AA~and the total energies are converged within $10^{-3}$ eV per unit cell. All other simulation parameters are same as that in section \ref{gamma}. A convergence with respect to $k$-grid densities for the smallest heterostructure of PdO$_2$-PdS$_2$ is shown in supplementary Figure S14. The heterostructures of the oxides on the dichalcogenides are generated using the $Hetero$2D workflow package\cite{boland2022computational}, with the constraints that the lattice mismatch is less than 10\% and coincident-site lattice area less than 200 \AA$^{2}$. Since the oxide monolayers are more flexible than the bulk dichalcogenides, for each heterostructure, the monolayer oxide is strained when placed on the dichalcogenide substrate. All the MO$_x$-MS$_2$ heterostructures for our focus materials are shown in Figure \ref{heterostructs}, while those of MoO$_3$-MoSe$_2$ and WO$_3$-WSe$_2$ are shown in supplementary Figure S2(a)-(b). 

Figure \ref{gammasl_fig}(c) shows the values of $\gamma_{S-L}$ for all the focus materials. We observe that $\gamma_{S-L}$ is positive, suggesting that the energy of the heterostructure exceeds the combined energy of the relaxed oxide monolayer and the dichalcogenide substrate. These positive values of $\gamma_{S-L}$ indicate that $\Gamma_{S-L}$ provides a driving force that enables delamination of the oxide monolayer and formation of the nanoscroll. 
 
\subsubsection{\textit{Ab initio} calculation of bending stiffness} \label{D}

\begin{figure*}
    \centering
    \includegraphics{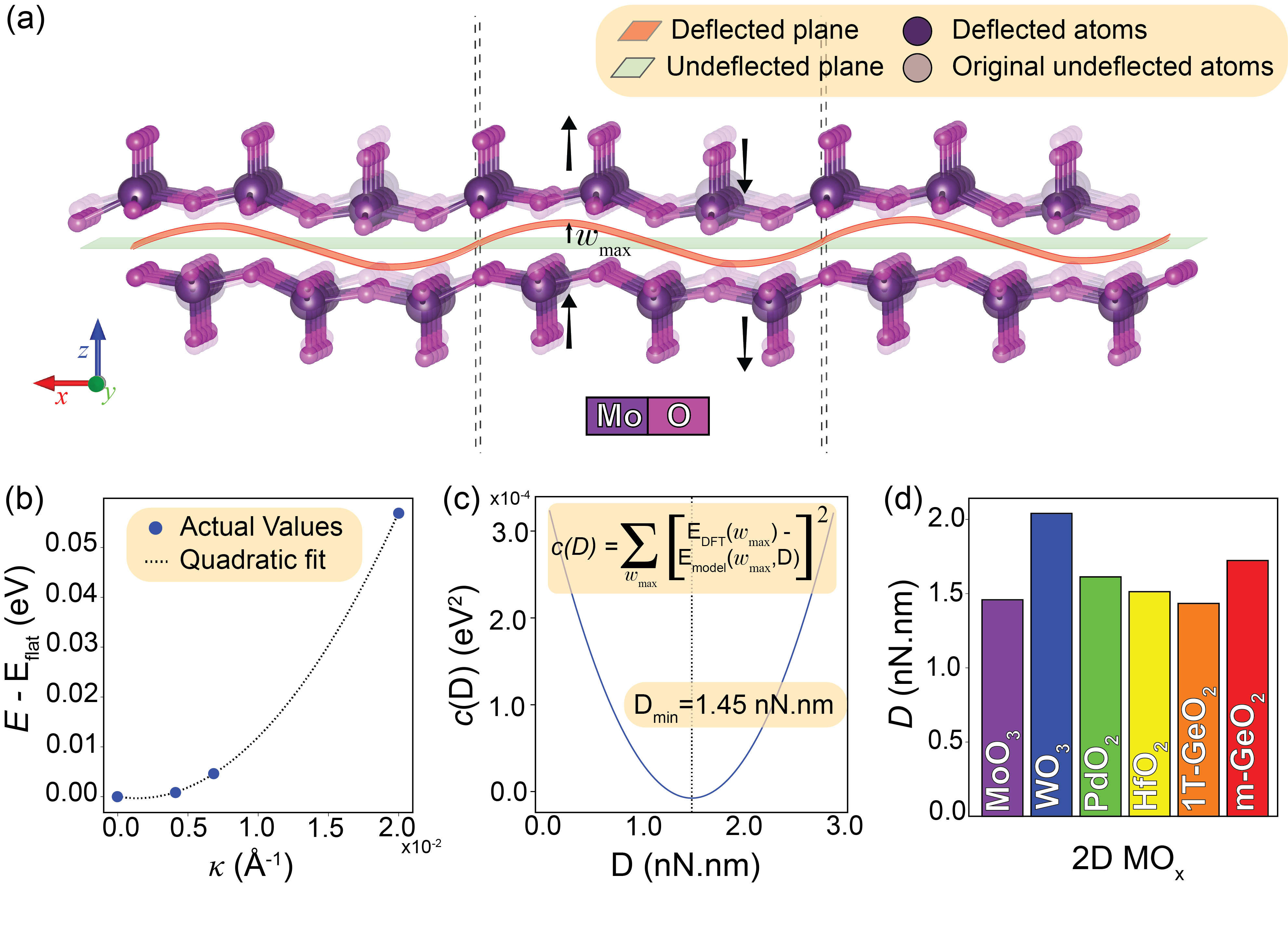}
    \caption{(a) A MoO$_3$ unitcell with applied curvature for the bending stiffness simulation. (b) The total energy of MoO$_3$ as a function of various curvatures, $\kappa$. (c)) The plot of the cost function, $c(D)$, as a function of the bending stiffness, $D$. (d) The bending stiffness, $D$, values for all the oxide monolayers.}
    \label{bending_stiff}
\end{figure*}
We adopt the method of Shirazian \textit{et al}\cite{Shirazian2022} to compute the bending stiffness, $D$, of the oxide monolayers. In this method, a curvature is applied to the monolayer unit cells such that symmetric perturbations are imposed on atoms in both the positive and negative $z$-directions, see Figure \ref{bending_stiff}(a). 

Shirazian \textit{et al}\cite{Shirazian2022} have shown that the values of $D$ can be obtained by comparing the DFT computed energies of the deformed 2D sheets with the energies of these sheets as estimated from beam theory. From  beam theory, the energies of curved 2D sheets are, 
\begin{equation}\label{emodel}
    E_{\text{model}} = \Big(\frac{6b}{L^3}\Big)D w_{\text{max}}^2
\end{equation}
where $D$, $b$, $L$, and $w_{\text{max}}$ are the bending stiffness, the sheet width, the sheet length, and the maximum deflection of the nanosheet, respectively. The maximum deflection, $w_{\text{max}}$, of a unit cell of length $L$, is given by
\begin{equation}
    w_{\text{max}} = \frac{1}{\kappa} - \sqrt{\frac{1}{\kappa^2} - \frac{L^2}{16}}
\end{equation}
where $\kappa$ is the curvature imposed on the nanosheet. Note that the value of $w_{\text{max}}$  is different for each unique curvature of the 2D sheet. The $w_{\text{max}}$ for  MoO$_3$ for an applied curvature is shown in Figure \ref{bending_stiff}(a). 

To estimate the values of $D$, the minima of a cost function that is based on the difference between DFT computed and beam theory energies can be utilized. This cost function is defined as,  
\begin{equation}
    c(D) = \sum_{w_{\text{max}}} (E_{\text{DFT}}(w_{\text{max}})-E_{\text{model}}(w_{\text{max}},D))^2
\end{equation}
where $E_{\text{DFT}}(w_{\text{max}})$ and $E_{\text{model}}(w_{\text{max}},D)$ are the DFT computed and beam theory energies respectively, of the deformed nanosheet corresponding to a maximum deflection $w_{\text{max}}$.

Figure \ref{bending_stiff}(c) shows the cost function as a function of $D$ for MoO$_3$. The DFT energies of the MoO$_3$ as a function of curvature are shown in Figure \ref{bending_stiff}(b). The bending stiffness of the remaining oxides considered in this study is shown in Figure \ref{bending_stiff}(d). The minimisation of the cost function for all the 2D oxides can be found in the supplementary Figures S4-S8.

For the bending stiffness DFT calculations, we used the Plane-Wave Self-Consistent Field (PWscf) package within the Quantum ESPRESSO distribution.\cite{qe1,qe2} Psuedopotentials were taken from the SSSP PBE Efficiency v1.2 library.\cite{prandini2018precision} The deformed structures as described earlier, for each oxide, are prepared for 3 unique curvatures $\kappa = $ $20\times10^{-3}$ \AA$^{-1}$, $6.67\times10^{-3}$ \AA$^{-1}$, $5\times10^{-3}$ \AA$^{-1}$(corresponding to radii of curvatures 50 \AA, 150 \AA, 250 \AA), as well as a flat sheet. The vacuum spacing in the $z$-direction is set to at least 30 \AA. We perform static calculation where the total energies in the single SCF step are converged to within $10^{-8}$ Ry. The Coulomb interaction between the periodic replicas in the out-of-plane direction is truncated using the method by Sohier \textit{et al}.\cite{sohier2017density}. For all the calculations, in order to capture the small energy variations across the structures with small changes in the curvatures, we select an extremely fine $k$-grid of $40 \times 40 \times 1$. This translates to a reciprocal density of about $5 \times 10^4$ \AA$^{-3}$ for the smallest unitcell. We also set extremely high plane wave cutoff of 100 Ry to obtain the accuracies needed for the estimation of D. 

Note that we compared the bending stiffness's obtained from this method with existing literature for two well studied 2D materials--graphene and MoS$_2$. The DFT computed energies and the minimized cost function plots for graphene and MoS$_2$ are shown in supplementary Figure S3. We computed the bending stiffness of graphene and MoS$_2$ to be 0.27 nN.nm and 2.31 nN.nm, respectively, which is in close agreement with the previously determined experimental and theoretical ranges of 0.19-0.27 nN.nm and 1.05-2.14 nN.nm for graphene and MoS$_2$, respectively.\cite{graphenebending1,graphenebending2,graphenebending3,mos2bending1,mos2bending2,mos2bending3}

\subsection{Nanoscroll Structure and Stability--Model Validation and Application} \label{NSresults}
\begin{figure}[h]
    \centering
    \includegraphics[width=5in]{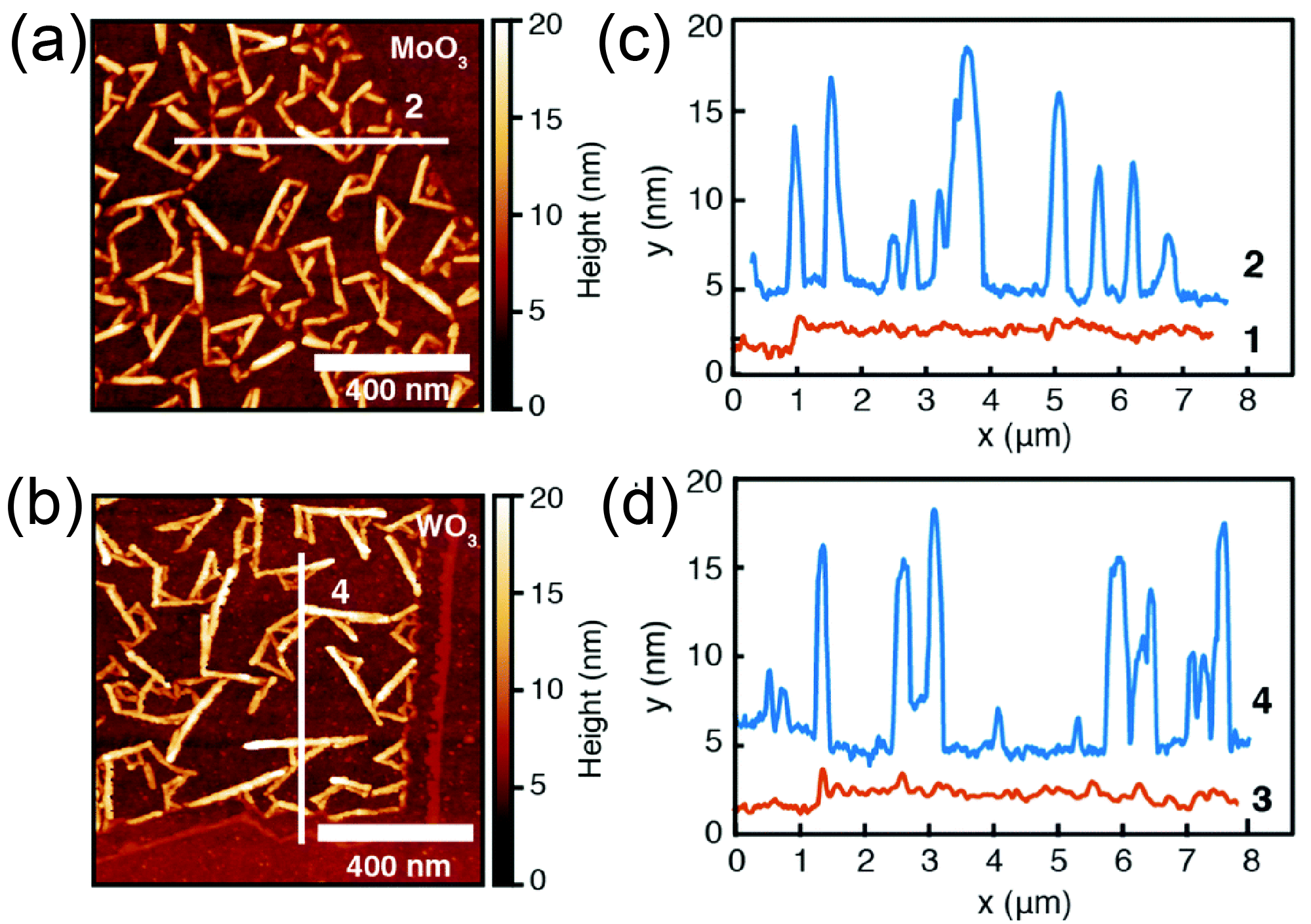}
    \caption{(a) and (b) show the surface AFM images post the PACS process on MoS$_2$ and WS$_2$ respectively. Oxide nanoscroll structures can clearly be seen. (c) and (d) show the height profiles along the lines 2 and 4 indicated in (a) and (b) respectively. Reproduced from Ref. \cite{Chu2017} with permission from the Royal Society of Chemistry}
    \label{height_profile}
\end{figure}

\begin{figure*}[ht]
    \centering
    \includegraphics{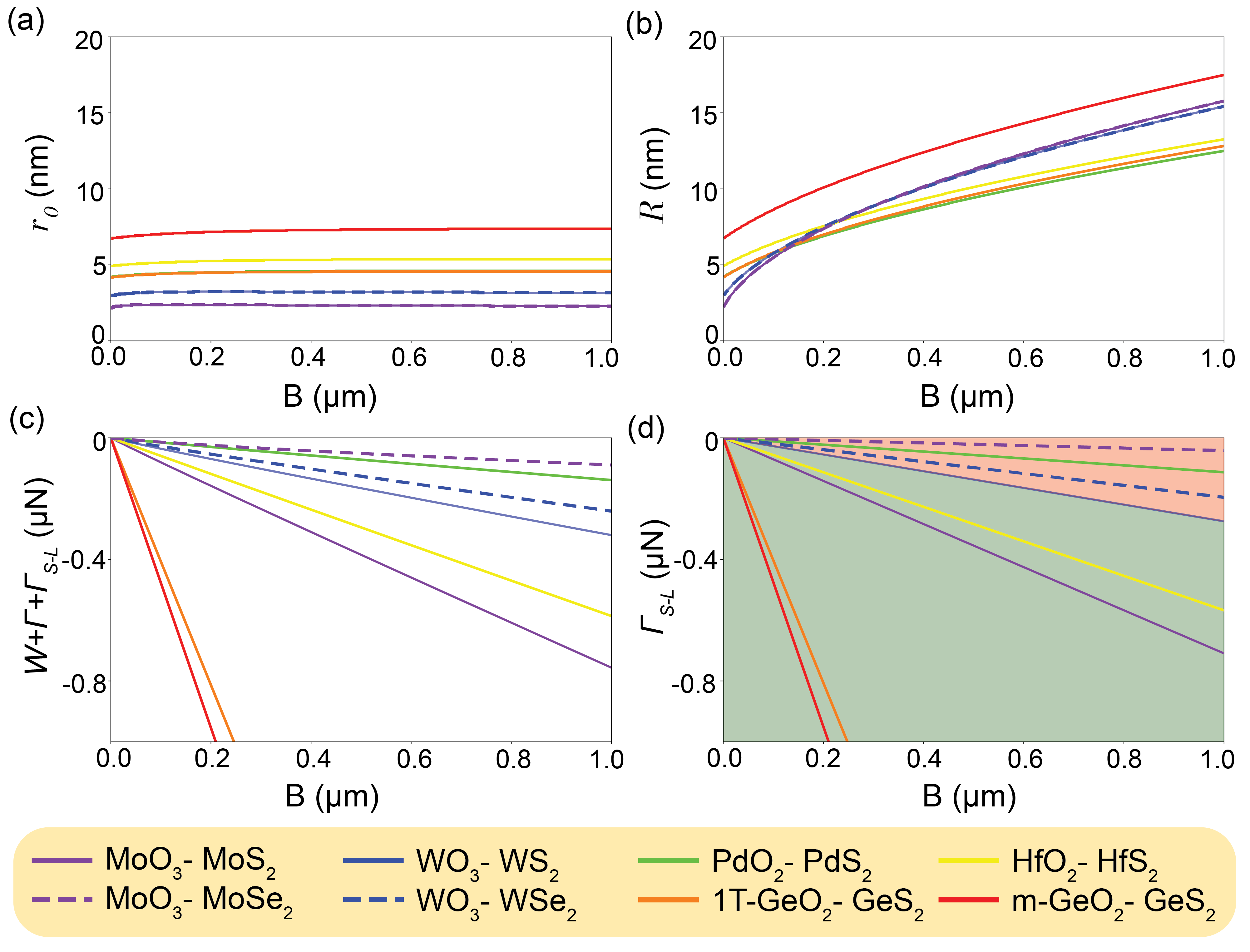}
    \caption{Shows the variation of (a) core radius at equilibrium (b) outer radius at equilibrium (c) total energy at equilibrium, and (d) layer-substrate interaction energy at equilibrium, when the length of the original nanosheet is varied from 0 to 1000 nm, for all our selected materials}
    \label{B_depend}
\end{figure*}

To validate our physio-mechanical model that is parameterized with vdW-corrected DFT simulations, we first examine its ability to explain the experimental observations of Chu \textit{et al.}. As mentioned above, Chu \textit{et al.} have shown that the PACS process applied to MoS$_2$ and WS$_2$ results in the formation of oxide nanoscrolls, but no nanoscrolls were formed when the PACS process was applied to MoSe$_2$ and WSe$_2$ precursors. In addition, their nanoscroll structure geometry can be established from their AFM images. Figure \ref{height_profile} shows that the MoS$_2$ and WS$_2$ precursor nanosheets had a width of about 200 nm and the AFM data indicate that the nanoscroll heights (i.e. their outer diameter) were about 15 nm. Our physio-mechanical model yields an outer diameter of 14.74 and 14.98 for MoS$_2$ and WS$_2$, respectively, for a sheet width of 200 nm. Thus the physio-mechanical model has an excellent agreement with the experimental measurements of the geometry of the scroll.

Figure \ref{B_depend}(a) shows the core radius of all the focus materials considered in this study. We find that the equilibrium core radius is nearly invariant of the sheet width. Figure \ref{B_depend}(b) shows the outer radii for the materials. The outer radii increase with increasing sheet width. We can also see that the core and outer radius for MoO$_3$-MoS$_2$, and MoO$_3$-MoSe$_2$ are nearly identical across the entire range of $B$. Similarly, the core and outer radius for WO$_3$-WS$_2$ and WO$_3$-WSe$_2$ are nearly identical across the entire range of $B$. Supplementary Table S3 shows that their $D$ and $\gamma$ values are the same; however, they exhibit different $\gamma_{S-L}$ values. Thus, it appears that the  $\gamma_{S-L}$, and in turn the layer substrate interaction $\Gamma_{S-L}$, has a negligible effect on the equilibrium scroll structure. The equilibrium scroll structure is, instead, determined only by the bending and interlayer surface energies.

In order to gain insight about the experimentally observed nanoscroll formation for MoS$_2$ and WS$_2$ precursors and not for the MoSe$_2$ and WSe$_2$ precursors, we plot the variation of the total energy, $E_{\mathrm{total}}=W+\Gamma+\Gamma_{S-L}$, in Figure \ref{B_depend}(c), as well as only the layer-substrate interaction, $\Gamma_{S-L}$, in Figure \ref{B_depend}(d), at equilibrium for $B$ ranging from 0 to 1000 nm. We can see that MoO$_3$-MoSe$_2$ has a less negative total energy value compared to MoS$_2$, which indicates a lower thermodynamic stability. The same trend is observed for WO$_3$-WSe$_2$. Thus, while the physio-mechanical model predicts that the Mo- and W- sulphide and selenides will yield nearly identical nanoscroll structures, the energies of the nanoscrolls obtained from the sulphides is lower, and thus they are more likely to form. 

Figure \ref{B_depend}(c) and (d) also plot the data for the remaining focus materials. For all of the materials, we find that the total energy reduces with increasing widths of the sheet, implying that nanoscrolls formed from wider sheets are more stable. Furthermore, we can see that the layer-substrate interaction energy, $\Gamma_{S-L}$, follows the total energy almost exactly. In fact, from Supplementary Figure S14, we see that for all the oxide-dichalcogenide pairs, the bending, $W$, and interlayer surface energies, $\Gamma$, almost cancel each other out, and the layer-substrate interaction energy, $\Gamma_{S-L}$ dominates the total energy. Thus, a study of only the $\gamma_{S-L}$, which are the slopes of the plots in Figure \ref{B_depend}(d), may be sufficient to predict the stability of equilibrium scroll structures for other precursors. Finally, since WO$_3$-WS$_2$ has been experimentally synthesized, we can adopt its $\gamma_{S-L}$ value as a benchmark for the successful generation of nanoscrolls through the PACS process. In other words, an oxide-dichalcogenide with $\gamma_{S-L}<\gamma_{S-L}^{\text{WO$_3$-WS$_2$}}$ has the thermodynamic driving force to yield nanoscrolls in the PACS process. Note that  MoO$_3$-MoS$_2$, satisfies this condition. For all $\gamma_{S-L}>\gamma_{S-L}^{\text{WO$_3$-WS$_2$}}$, such as in MoO$_3$-MoSe$_2$, WO$_3$-WSe$_2$, and PdO$_2$-PdS$_2$ there isn't enough total energy to stabilize the nanoscroll structure. Thus Figure \ref{B_depend}(d) can thus be separated into a stable (green shaded region) and unstable (red shaded region) nanoscroll energy regions. Based on this energy criteria, nanoscrolls of 1T-GeO$_2$-GeS$_2$, m-GeO$_2$-GeS$_2$, and HfO$_2$-HfS$_2$ can be obtained while that of PdO$_2$-PdS$_2$ may not be thermodynamically stable. 

\section{Conclusions}
In summary, we shed significant light on the thermodynamics of nanoscroll structure formation, and especially on non vdW bonded oxides formed from dichalcogenide precursors by plasma-assisted conversion. Motivated by the objective of studying the formation of non vdW bonded nanoscroll structures, we began our study by screening  materials from known materials databases to identify potential candidates that can act as precursors for the PACS process, a recipe that is known to produce non vdW bonded oxide nanoscrolls. After the selection of dichalcogenide candidates and their corresponding oxides, we developed a physio-mechanical model to discuss the stability of nanoscroll structures. In this model we accounted for the often ignored role of the substrate on nanoscroll stability. We then used \textit{ab initio} vdW-corrected DFT calculations to determine the constituent physical parameters of the model such as the bending stiffness, $D$, interlayer energies per unit area, $\gamma$, layer-substrate energies per unit area, $\gamma_{S-L}$, and interlayer spacing, $h$. 

We compared our physio-mechanical model predictions with those observed in previously reported experiments and found that we have an exact match for the structural parameters of the nanoscrolls. We also find that while the bending and interlayer surface energies effectively determine the stable nanoscroll structure, the layer-substrate interaction energies have a minimal effect on the stable structure. Instead, we find that the layer-substrate interaction energies determine the absolute stability of the stable structure. Finally, using our determined values of the layer-substrate interactions combined with the experimental findings, we find a qualitative metric that divides the phase space of nanoscrolls formed atop substrates into two regions of stability and instability.  

Our work thus greatly expands our understanding of nanoscroll formation by refining existing physio-mechanical models. We achieved this by incorporating the previously overlooked influence of substrates and utilizing \textit{ab initio} computed parameters. This work lays the foundation for systematic, predictive 
and mechanistic studies of oxide nanoscroll formation from dichalcogenide precursors, facilitating their cost-effective and large-scale experimental implementation and applications.

\medskip
\textbf{Supporting Information} \par 
Supporting Information is available from the Wiley Online Library or from the author.

\medskip
\textbf{Acknowledgements} \par 
The authors acknowledge support by NSF DMR under Grant No. DMR-1906030 and start-up funds from Arizona State University, USA. The authors also acknowledge the San Diego Supercomputer Center under the NSF-XSEDE Award No. DMR150006 and the Research Computing at Arizona State University for providing HPC resources. This research used resources of the National Energy Research Scientific Computing Center, a DOE Office of Science User Facility supported by the Office of Science of the U.S. Department of Energy under Contract No. DE-AC02-05CH11231. 

\medskip

%
\bibliographystyle{MSP}
\bibliography{references.bib}

\begin{thebibliography}{10}
\providecommand{\url}[1]{\texttt{#1}}
\providecommand{\urlprefix}{URL }

\bibitem{Chu2017}
X.~S. Chu, D.~O. Li, A.~A. Green, Q.~H. Wang,
\newblock \emph{Journal of Materials Chemistry C} \textbf{2017}, \emph{5}
  11301.

\bibitem{tmdctmo}
G.~{\"O}zbal, R.~T. Senger, C.~Sevik, H.~Sevin{\c{c}}li,
\newblock \emph{Physical Review B} \textbf{2019}, \emph{100}, 8 085415.

\bibitem{tmocatalysis1}
D.-H. Seo, A.~Urban, G.~Ceder, et~al.,
\newblock \emph{Physical Review B} \textbf{2015}, \emph{92}, 11 115118.

\bibitem{tmocatalysis2}
V.~C. Kolluru, R.~G. Hennig,
\newblock \emph{Physical Review Materials} \textbf{2020}, \emph{4}, 4 045803.

\bibitem{cnshydrogen}
V.~Coluci, S.~Braga, R.~Baughman, D.~Galvao,
\newblock \emph{Physical Review B} \textbf{2007}, \emph{75}, 12 125404.

\bibitem{Mos2nsphoto}
X.~Fang, P.~Wei, L.~Wang, X.~Wang, B.~Chen, Q.~He, Q.~Yue, J.~Zhang, W.~Zhao,
  J.~Wang, et~al.,
\newblock \emph{ACS applied materials \& interfaces} \textbf{2018}, \emph{10},
  15 13011.

\bibitem{Sayyad2023}
M.~Sayyad, Y.~Qin, J.~Kopaczek, A.~Gupta, N.~Patoary, S.~Sinha, E.~Benard,
  A.~Davis, K.~Yumigeta, C.~L. Wu, H.~Li, S.~Yang, I.~S. Esqueda, A.~Singh,
  S.~Tongay,
\newblock \emph{Advanced Functional Materials} \textbf{2023}, \emph{33}.

\bibitem{gupta2022anomalous}
A.~Gupta, T.~Biswas, A.~Singh,
\newblock \emph{Journal of Applied Physics} \textbf{2022}, \emph{132}, 24.

\bibitem{guo2021exfoliation}
Y.~Guo, A.~Gupta, M.~S. Gilliam, A.~Debnath, A.~Yousaf, S.~Saha, M.~D. Levin,
  A.~A. Green, A.~K. Singh, Q.~H. Wang,
\newblock \emph{Nanoscale} \textbf{2021}, \emph{13}, 3 1652.

\bibitem{li2019anomalous}
H.~Li, K.~Wu, S.~Yang, T.~Boland, B.~Chen, A.~K. Singh, S.~Tongay,
\newblock \emph{Nanoscale} \textbf{2019}, \emph{11}, 42 20245.

\bibitem{haastrup2018computational}
S.~Haastrup, M.~Strange, M.~Pandey, T.~Deilmann, P.~S. Schmidt, N.~F. Hinsche,
  M.~N. Gjerding, D.~Torelli, P.~M. Larsen, A.~C. Riis-Jensen, et~al.,
\newblock \emph{2D Materials} \textbf{2018}, \emph{5}, 4 042002.

\bibitem{Negreiros2019}
F.~R. Negreiros, T.~Obermüller, M.~Blatnik, M.~Mohammadi, A.~Fortunelli, F.~P.
  Netzer, S.~Surnev,
\newblock \emph{The Journal of Physical Chemistry C} \textbf{2019}, \emph{123}
  27584.

\bibitem{weng2018honeycomb}
J.~Weng, S.-P. Gao,
\newblock \emph{Physical Chemistry Chemical Physics} \textbf{2018}, \emph{20},
  41 26453.

\bibitem{Singh2017}
A.~K. Singh, B.~C. Revard, R.~Ramanathan, M.~Ashton, F.~Tavazza, R.~G. Hennig,
\newblock \emph{Physical Review B} \textbf{2017}, \emph{95} 155426.

\bibitem{Shi2010}
X.~Shi, N.~M. Pugno, H.~Gao,
\newblock \emph{Journal of Computational and Theoretical Nanoscience}
  \textbf{2010}, \emph{7} 517.

\bibitem{Shi2010_2}
X.~Shi, N.~M. Pugno, H.~Gao,
\newblock \emph{China Acta Mechanica Solida Sinica} \textbf{2010}, \emph{23}.

\bibitem{jain2013commentary}
A.~Jain, S.~P. Ong, G.~Hautier, W.~Chen, W.~D. Richards, S.~Dacek, S.~Cholia,
  D.~Gunter, D.~Skinner, G.~Ceder, et~al.,
\newblock \emph{APL materials} \textbf{2013}, \emph{1}, 1.

\bibitem{GeS2synthesis}
L.~Kulikova, L.~Lityagina, I.~Zibrov, T.~Dyuzheva, N.~Nikolaev, V.~Brazhkin,
\newblock \emph{Inorganic Materials} \textbf{2014}, \emph{50} 768.

\bibitem{ong2013python}
S.~P. Ong, W.~D. Richards, A.~Jain, G.~Hautier, M.~Kocher, S.~Cholia,
  D.~Gunter, V.~L. Chevrier, K.~A. Persson, G.~Ceder,
\newblock \emph{Computational Materials Science} \textbf{2013}, \emph{68} 314.

\bibitem{icsd}
D.~Zagorac, H.~M{\"u}ller, S.~Ruehl, J.~Zagorac, S.~Rehme,
\newblock \emph{Journal of applied crystallography} \textbf{2019}, \emph{52}, 5
  918.

\bibitem{Kresse1}
G.~Kresse, J.~Hafner,
\newblock \emph{Physical Review B} \textbf{1993}, \emph{47} 558.

\bibitem{Kresse2}
G.~Kresse, J.~Hafner,
\newblock \emph{Physical Review B} \textbf{1994}, \emph{49}, 20 14251.

\bibitem{Kresse3}
G.~Kresse, J.~Furthm\"{u}ller,
\newblock \emph{Computational Materials Science} \textbf{1996}, \emph{6} 15.

\bibitem{Kresse4}
G.~Kresse, J.~Furthm{\"{u}}ller,
\newblock \emph{Physical Review B} \textbf{1996}, \emph{54} 11169.

\bibitem{Kresse5}
G.~Kresse, D.~Joubert,
\newblock \emph{Physical Review B} \textbf{1999}, \emph{59} 1758.

\bibitem{Klimes2011}
J.~Klimeš, D.~R. Bowler, A.~Michaelides,
\newblock \emph{Physical Review B} \textbf{2011}, \emph{83}, 19 195131.

\bibitem{boland2022computational}
T.~M. Boland, A.~K. Singh,
\newblock \emph{Computational Materials Science} \textbf{2022}, \emph{207}
  111238.

\bibitem{Shirazian2022}
F.~Shirazian, R.~A. Sauer \textbf{2022}.

\bibitem{qe1}
P.~Giannozzi, S.~Baroni, N.~Bonini, M.~Calandra, R.~Car, C.~Cavazzoni,
  D.~Ceresoli, G.~L. Chiarotti, M.~Cococcioni, I.~Dabo, et~al.,
\newblock \emph{Journal of physics: Condensed matter} \textbf{2009}, \emph{21},
  39 395502.

\bibitem{qe2}
P.~Giannozzi, O.~Andreussi, T.~Brumme, O.~Bunau, M.~B. Nardelli, M.~Calandra,
  R.~Car, C.~Cavazzoni, D.~Ceresoli, M.~Cococcioni, et~al.,
\newblock \emph{Journal of physics: Condensed matter} \textbf{2017}, \emph{29},
  46 465901.

\bibitem{prandini2018precision}
G.~Prandini, A.~Marrazzo, I.~E. Castelli, N.~Mounet, N.~Marzari,
\newblock \emph{npj Computational Materials} \textbf{2018}, \emph{4}, 1 72,
  \href{http://materialscloud.org/sssp}{http://materialscloud.org/sssp}.

\bibitem{sohier2017density}
T.~Sohier, M.~Calandra, F.~Mauri,
\newblock \emph{Physical Review B} \textbf{2017}, \emph{96}, 7 075448.

\bibitem{graphenebending1}
Y.~Wei, B.~Wang, J.~Wu, R.~Yang, M.~L. Dunn,
\newblock \emph{Nano letters} \textbf{2013}, \emph{13}, 1 26.

\bibitem{graphenebending2}
K.~N. Kudin, G.~E. Scuseria, B.~I. Yakobson,
\newblock \emph{Physical Review B} \textbf{2001}, \emph{64}, 23 235406.

\bibitem{graphenebending3}
E.~Han, J.~Yu, E.~Annevelink, J.~Son, D.~A. Kang, K.~Watanabe, T.~Taniguchi,
  E.~Ertekin, P.~Y. Huang, A.~M. van~der Zande,
\newblock \emph{Nature materials} \textbf{2020}, \emph{19}, 3 305.

\bibitem{mos2bending1}
S.~Xiong, G.~Cao,
\newblock \emph{Nanotechnology} \textbf{2016}, \emph{27}, 10 105701.

\bibitem{mos2bending2}
J.~W. Jiang,
\newblock \emph{Nanotechnology} \textbf{2014}, \emph{25}, 35 355402.

\bibitem{mos2bending3}
G.~Casillas, U.~Santiago, H.~Barro\'n, D.~Alducin, A.~Ponce,
  M.~Jose\''-Yacama\'n,
\newblock \emph{The Journal of Physical Chemistry C} \textbf{2015}, \emph{119},
  1 710.

\end{thebibliography}



\end{document}